# CCT5: A Code-Change-Oriented Pre-Trained Model

Bo Lin
linbo19@nudt.edu.cn
College of Computer Science,
National University of Defense
Technology
Changsha, China

Shangwen Wang
wangshangwen13@nudt.edu.cn
College of Computer Science,
National University of Defense
Technology
Changsha, China

Zhongxin Liu
liu_zx@zju.edu.cn
Zhejiang University
Hangzhou, China

Yepang Liu
liuyp1@sustech.edu.cn
Department of Computer Science and
Engineering, Southern University of
Science and Technology
Shenzhen, China

Xin Xia
xin.xia@acm.org
Zhejiang University
Hangzhou, China

Xiaoguang Mao
xgmao@nudt.edu.cn
College of Computer Science,
National University of Defense
Technology
Changsha, China

## ABSTRACT

Software is constantly changing, requiring developers to perform several derived tasks in a timely manner, such as writing a description for the intention of the code change, or identifying the defect-prone code changes. Considering that the cost of dealing with these tasks can account for a large proportion (typically around 70 percent) of the total development expenditure, automating such processes will significantly lighten the burdens of developers. To achieve such a target, existing approaches mainly rely on training deep learning models from scratch or fine-tuning existing pre-trained models on such tasks, both of which have weaknesses. Specifically, the former uses comparatively small-scale labelled data for training, making it difficult to learn and exploit the domain knowledge of programming language hidden in the large-amount unlabelled code in the wild; the latter is hard to fully leverage the learned knowledge of the pre-trained model, as existing pre-trained models are designed to encode a single code snippet rather than a code change (i.e., the difference between two code snippets). We propose to pre-train a model specially designed for code changes to better support developers in software maintenance. To this end, we first collect a large-scale dataset containing 1.5M+ pairwise data of code changes and commit messages. Based on these data, we curate five different tasks for pre-training, which equip the model with diverse domain knowledge about code changes. We fine-tune the pre-trained model, CCT5, on three widely-studied tasks incurred by code changes and two tasks specific to the code review process. Results show that CCT5 outperforms both conventional deep learning approaches and existing pre-trained models on these tasks.

## 1 INTRODUCTION

Software undergoes continuous changes during the maintenance phase to fix defects, change execution logic, make the processing more efficient, or introduce new features [14, 54, 57, 64]. Because of the omnipresence of code changes, developers need to deal with a number of derived tasks (referred to as **code-change-related tasks** in this paper). For instance, the interview conducted by Fritz and Murphy [15] showed that developers have to frequently answer the question "Why were the code changes introduced?" in their daily development tasks. After comprehending the intention of code change, they may further need to estimate the impact of software changes [52], troubleshoot unexpected behavior [28], monitor the maintenance of code clones [49], or update the associated code comments [35]. Statistics have shown that maintaining software systems demands possibly as high as 70 percent of the total development efforts [37]. Furthermore, a user study indicates that there is an urgent need of tool support for code-change-related tasks, such as commit message generation and defect prediction [61].

Prior works proposed to leverage deep learning (DL) techniques to deal with code-change-related tasks and have achieved promising results [35, 43]. Currently, there are typically two ways to apply deep learning techniques. The first is to train deep learning models from scratch with a comparatively small-scale, manually-labelled, and task-specific dataset (referred to as *non-pre-training* in this paper) [23, 36, 50, 53]. Despite the progress such approaches have achieved, their effectiveness sometimes cannot be as promising as expected (possibly because of the lack of training data [35]), and there is still a large room for improvement [34, 41, 70]. For instance, Liu *et al.* [41] found that, when generating commit messages that describe the intention of code changes, a simple heuristic can outperform a sophisticated DL model and save a lot of computing resources at the same time. Another way is to leverage the state-of-the-art *pre-training* paradigm, where the models are first pre-trained on large-scale unlabelled datasets to be equipped with domain knowledge of programming language (PL) and natural language (NL), and then fine-tuned on various downstream tasks [1, 13, 18, 72]. By doing so, the parameters of the trained model can store some common knowledge compared with random initialization. Specifically, one can fine-tune existing pre-trained models on code-change-related tasks [33, 79]. However, existing pre-training techniques are mainly targeted at tasks related to encoding and understanding a given code snippet (referred to as **code-related tasks** in this paper), such as code search [17], code summarization [25], and defect detection [32]. Specifically, the designed pre-training tasks (e.g., the Masked Language Modeling in CodeBERT [13]) typically take a code snippet and its paired documentation (i.e., the comment) as inputs. Thus, the learned domain knowledge is generally related to the syntactic and semantic information of code snippets, which can hardly be exploited to encode and understand code changes. That is to say, adopting the second way will inevitably lead to inconsistent inputs and objectives between pre-training and fine-tuning since code-related and code-change-related tasks have natural differences: the former deals with a code snippet and the key challenge is to capture the syntactic and semantic information of a code snippet; while the latter deals with two code snippets and the key challenge is to understand the differences. Consequently, it is sub-optimal to fine-tune existing pre-trained code models for code-change-related tasks [67].

To help developers better deal with code changes and address the limitations mentioned above, our basic idea is that models can be equipped with different domain knowledge and thus be applied to different tasks if they are pre-trained by different tasks [8, 33, 72, 79]. Therefore, we propose a code-change-oriented pre-trained model, CCT5, which is built on top of the well-known Text-To-Text-Transfer Transformer (T5) model [56], but pre-trained with code-change-specific inputs and objectives. CCT5 mainly embodies two advantages: First, by adopting the pre-training paradigm, the domain knowledge of code changes hidden in a large amount of unlabelled data can be absorbed by the model. Second, by designing specific pre-training tasks for code changes, the domain knowledge learned by the pre-trained model can be easily transferred to code-change-related downstream tasks. To achieve our target, we first build a large-scale dataset (named **CodeChangeNet**) for pre-training. Specifically, we collect 1.5M+ pairwise data of code change and commit message from popular GitHub projects written in six widely-used programming languages. As the NL description of the code change during pre-training, the commit message plays a similar role to the code comment in existing pre-training techniques. After that, we design five different pre-training tasks for learning the domain knowledge about code changes, which take as inputs the differences between two code snippets (i.e., code diffs [33]), and enable the model to align the NL and PL representations, generate fluent NL descriptions and complete code snippets, and be aware of the program structure, respectively.

To evaluate the effectiveness of CCT5, we fine-tune it on three widely-studied downstream tasks, which are commit message generation [68], just-in-time comment update [43], and just-in-time defect prediction [50]. Results show that CCT5 outperforms both the state-of-the-art non-pre-training techniques and existing pre-trained models on the three tasks consistently. For instance, on a large-scale multi-linguistic benchmark for the commit message generation task, CCT5 outperforms NNGen (the state-of-the-art non-pre-training technique) and CodeT5 (the state-of-the-art pre-trained model) by 24% and 22%, respectively, in terms of the BLEU values. Besides, CCT5 can also generalize well to code review tasks: it outperforms CodeReviewer [33], a recently-proposed pre-training technique specially designed for the code review process, on the code change quality estimation and review generation tasks.

In summary, our study makes the following contributions:

- **Dataset:** We collect and release a large-scale dataset for performing code-change-oriented pre-training, i.e., CodeChangeNet.
- **CCT5:** We propose totally five carefully-curated pre-training tasks, based on which we release the first code-change-oriented pre-trained model, i.e., CCT5.
- **Performance assessment:** We perform extensive experiments to assess the performance of CCT5. CCT5 achieves the state-of-the-art performance on three widely-studied code-change-related tasks and two tasks specific to the code review process.

## 2 BACKGROUND AND RELATED WORKS

### 2.1 Code Change and Its Related Tasks

Software is constantly changing as new features are added, bugs are fixed, and performance is enhanced [64]. Specifically, it has been shown that the Linux kernel changes 5.5 times per hour [7]. In another study, Jiang *et al.* found that there are over 2M code changes in the most popular 1K open-source repositories on GitHub[27]. Both the above studies demonstrate the widespread existence of code changes. Along with the changes of code, developers need to solve many derived tasks. For instance, a developer may need to write a description for the code changes he/she made for better communication among the development team [7, 27]; or in some other cases, he/she may need to check if a code change made by others will induce program defects or not [29, 60]. Given that, it is not surprising that the cost of program maintenance can reach around 70% of the total expenditure, as reported by Lehman [31]. Therefore, techniques that can automate such code-change-related tasks hold great potential to boost developers' productivity.

### 2.2 Code Change Representation Techniques

Previously studies have demonstrated that compared with conventional heuristic-based approaches, deep learning techniques can perform better on code-change-related tasks [36, 39, 43, 69]. Consequently, to deal with code-change-related tasks, recent studies focus on learning the distributed representations of code changes using DL techniques.

CC2Vec [23] adopts a hierarchical attention network to successively build vector representations for code lines, code chunks, and finally the entire code change. It achieves state-of-the-art performance on a number of code-change-related tasks such as just-in-time defect prediction [29]. Yin *et al.* [77] experimented with two different ways to embed a code change: treating it as a code token sequence and treating it as the difference between two Abstract Syntax Trees (ASTs). The results show that the representations could be better if the structure-level information is involved. Recent studies also utilize the *AST path* technique [3] for representing the code change [35, 36], where the AST is split into changed and unchanged parts and the paths connecting different leaf nodes are embedded to represent each part. Despite the progress achieved by these techniques, they are all trained from scratch with a comparatively small-scale labelled dataset. It is hard for them to learn and use the domain knowledge of programming language hidden in the large amount of unlabelled code in the wild, which limits their effectiveness [35].

### 2.3 Existing Pre-Training Techniques

Training a deep learning model from scratch requires a large amount of labelled data, which is rather labor-intensive. To reduce the burden of manual labelling, pre-training techniques have been proposed recently, with the aim of equipping the model with commensense knowledge using unlabelled data, after which the model can be fine-tuned on downstream tasks with relatively small-scale labelled data. Such a paradigm was first proposed in the Natural Language Processing (NLP) domain [9, 55], and then adapted to code-related tasks by designing pre-training tasks to learn the domain knowledge of programming language or build the connections between a code snippet and its associated documentation [13, 18, 72].

Existing pre-trained models mainly target two types of code-related tasks: generation and understanding [13, 78]. The generation tasks denote those that require the generation of a sequence of tokens/words either in program languages (e.g., code generation [8, 76] and code repair [65]) or natural languages (e.g., code summarization [25, 30]). The understanding tasks denote those that require to produce vectorized representations of programs and then perform downstream tasks such as classification and ranking (e.g., code search [17, 66], code clone detection [71], and defect detection [81]). Although existing pre-training techniques such as CodeBERT [13], GraphCodeBERT [18], and CodeT5 [72] outperform traditional supervised learning techniques for the above tasks [45], few of them target code-change-related tasks. As introduced, the pre-training of these models focuses on capturing the syntactic and semantic information of a code snippet, whereas dealing with code-change-related tasks requires the model to encode and understand the differences between two code snippets (rather than a single code snippet). Consequently, the inputs to the models will be different from those used for pre-training when the models are fine-tuned for code-change-related downstream tasks. Such inconsistency makes the knowledge learned by pre-trained models hard to be fully exploited, easily leading to sub-optimal results for downstream tasks [40, 67]. In the literature, CoditT5 [79] is evaluated on

```
📄 Enable subqueries in gsheetsdb
📄 superset/db_engine_specs/gsheets.py
25  25     engine = "gsheets"
26      -  allows_subqueries = False
    26 +  allows_subqueries = True
```

**Figure 1: An example of code change with commit message.**

a code-change-related downstream task (i.e., just-in-time comment update). However, its pre-training paradigm is designed for better editing the code and comment rather than capturing the syntactic and semantic information of code changes. To our best knowledge, the only existing pre-trained model that aims at modelling code changes is CodeReviewer [33], but it is specially designed for code review tasks (e.g., generating the review comment and refining the code based on the review comment). Beyond such tasks, there are many other code-change-related tasks where an automated tool can significantly lighten the burdens on developers, such as commit message generation [41], code comment update [43], and defect prediction [53]. Thus, the literature lacks a pre-trained model that can perform well on diverse code-change-related tasks. Moreover, from the technical perspective, the pre-training of CodeReviewer ignores program structure information, which has been shown to be critical for the model's capability [18].

## 3 THE DATASET: CODECHANGENET

Pre-training techniques in the software engineering domain rely on capturing the syntactic and semantic information of code snippets. Previously, for code-related pre-trained models, the target is achieved by using the pairwise data, i.e., the code and its corresponding comment, for pre-training. The comment, which summarizes the main functionality of the code, provides the model with a way to understand the semantic information of the code. Similarly, if we are going to pre-train a model for code changes, we also need a way to reflect the semantic information of the code change. We note that while committing a code change to a version control system, developers need to document their changes using a commit message, which usually summarizes what happens in the change [7, 47]. A concrete example is shown in Figure 1. A developer changes the value of a variable from `false` to `true`, and the associated commit message describes this change as "enable subqueries". Therefore, in this work, we choose to use the commit messages to serve as the natural language descriptions of code changes. Existing datasets on commit messages, however, are usually small-scale. For instance, the dataset provided by Jiang *et al.* [27] only contains 32K commit messages. Such a scale cannot ensure the adequacy of pre-training (cf. the dataset used for code-related pre-training, i.e., the CodeSearchNet [26], contains around 2M code-comment pairs). Therefore, we propose to build a multi-linguistic dataset with large-scale pairwise data of code changes and commit messages for pre-training, named **CodeChangeNet**. In the following sections, we present the details of the building of dataset.

### 3.1 Data Collection

Nowadays, many projects are hosted on software development platforms such as GitHub. With developers continuously committing their code changes, there is a large amount of readily-available commit data including the exact content of code changes and commit messages associated with code changes. We thus build our CodeChangeNet dataset based on commit data collected from open-source projects in GitHub.

Table 1: Statistics of the pre-training dataset.

| Language | Projects | Commits | Data Size |
|---|---|---|---|
| Python | 9,627 | 519k | 16.0GB |
| Javascript | 11,925 | 97k | 3.2GB |
| Ruby | 1,839 | 62k | 0.5GB |
| Go | 4,110 | 185k | 4.0GB |
| Java | 5,645 | 461k | 13.0GB |
| PHP | 2,384 | 230k | 2.8GB |
| Total | 35,530 | 1,524k | 39.6GB |

[†] Projects of the dataset are accessed in October 2022.

Following CodeSearchNet, we collect projects in six popular programming languages, which are Go, Java, JavaScript, PHP, Python, and Ruby. To ensure the quality of our dataset, we collect data from projects with high popularity, which is indicated by the number of stars. Specifically, we select projects whose numbers of stars are more than 500 (such a criterion is widely-used by existing studies to indicate that a project is popular [4, 74]). We then remove projects that do not have a license or whose licenses do not explicitly permit the re-distribution of parts of the project.

To collect commit data, we use GitHub REST API to crawl project information from GitHub, following the common practice in the mining software repository (MSR) domain [21, 33]. Specifically, by calling GitHub API, the detailed repository information of each project such as the commit data can be accessed and stored in a JSON file. Such commit data includes code changes (i.e., the original file, new file, and the code diff) and commit messages. Finally, a set of pairs ($cc_i$, $cm_i$), where $cc_i$ is a code change and $cm_i$ is the associated commit message, is collected as the initial data of CodeChangeNet.

### 3.2 Data Filtering

To further ensure the data used for pre-training is of high quality, we perform the following preprocessing steps to filter low-quality data:

- Pairs whose $cm_i$ is shorter than three tokens (including three) are removed to ensure that the commit message is descriptive. This decision follows CodeSearchNet which restricts that the comment of the code should contain more than three tokens.
- Pairs whose $cc_i$ involves more than 100 tokens in the code diff are removed to ensure that the model will not be affected by such extremely complex code changes. This decision follows existing studies which build commit-related datasets [5, 27, 68].
- Pairs whose $cc_i$ occurs in the test files are removed since we focus on code changes in the source code. This decision also follows the preprocessing of CodeSearchNet.
- Pairs from those projects that have been used to build downstream tasks (which will be introduced later in Section 5) are removed to avoid data leakage.

The resulting CodeChangeNet dataset contains about 1.5M of <code change, commit message> pairs from 35,530 projects. Such a scale is comparable to that of CodeSearchNet and thus can ensure the adequacy of pre-training. The detailed statistics of the dataset are illustrated in Table 1.

## 4 CCT5

In this section, we introduce the details of our CCT5, including the model architecture, the input-output representations of the model, and the five different pre-training tasks designed for the model.

### 4.1 Model Architecture

Following the T5 model [56], CCT5 uses an encoder-decoder architecture. The encoder and decoder both have 12 Transformer layers and in each layer, 12 attention heads are used to perform the multi-head attention calculation, leading to the total parameter size being 220M. Such an architecture is widely-used by state-of-the-art pre-trained models [8, 16, 33, 72].

Following existing studies [8, 33], we initialize the parameters of CCT5 with the values from CodeT5, with the aim of equipping the model with some domain knowledge of the programming language. CCT5 is then further trained on five different pre-training tasks and then fine-tuned on various downstream tasks.

### 4.2 Input-Output Representation

Since CCT5 is designed to address code-change-related tasks, a fundamental problem is thus how to represent code changes. If we send the token sequences of the old code and new code into the model, the inputs would be extremely long since a code change may happen across different files, which could make the model hard to converge [2]. Instead, we use *code diff* to represent the code changes which is shown to be effective [33]. A diff file is generated by comparing the files before and after the code change. Specifically, there are one or more diff hunks in a diff file, and each diff hunk contains three different types of information: lines deleted in the change (indicated by a "-" at the beginning of each line, e.g., the red line in Figure 1), lines added in the change (indicated by a "+" at the beginning of each line, e.g., the green line in Figure 1), and surrounding lines unchanged in the change (which serve as the context information of the code change, e.g., the line 25 in Figure 1). One advantage of using the diff format is that it reduces the input length to a large extent, as the unchanged lines occur only once.

For downstream tasks, CCT5 takes the code diff as input and then performs a number of different tasks. Following the standard manner of the Transformer, the input is treated as a token sequence. We reuse the tokenizer from CodeT5 to split the code into a sequence of tokens. After that, a special token $[CLS]$ is prepended to the sequence, making the input with the form of { $[CLS]$, $c_1$, ..., $c_n$ }, where $c_i$ is a source code token and $n$ denotes the length of the code token sequence. This decision follows existing studies [13, 18, 72], and the token $[CLS]$ will be used as the representation of the code change in understanding tasks (which will be described in more detail in Section 5.3). To help the model better understand the code change, we also insert a special token in front of each line: for deleted lines, we insert the token $[DEL]$; for added lines, we insert the token $[ADD]$; and for unchanged lines, we insert the token $[KEEP]$. In the pre-training phase, some of our pre-training tasks take as inputs the code change and the commit message simultaneously, with the aim of building the semantic connection between the programming language and the natural language. For such tasks, the inputs will be { [CLS], $c_1, \ldots, c_n$, [MSG], $m_1, \ldots, m_l$ }, where [MSG] is a token separating the code tokens and commit message tokens, $m_i$ is a commit message token, and $l$ denotes the length of the commit message token sequence.

## 4.3 Pre-Training Tasks

An important goal of CCT5 is designed to accurately capture the semantic information of a given code change. To achieve so, we need to build the semantic connection between the code change and the commit message during the pre-training phase. We design totally four pre-training tasks to fulfill this target. Besides, inspired by GraphCodeBERT [18], we also design a pre-training task to make the model be aware of the program structure. Figure 2 gives an illustration of the five pre-training tasks of CCT5. Details about these tasks will be introduced below.

*4.3.1 Masked Language Modeling.* The Masked Language Modeling (MLM) pre-training task is widely used in previous studies [13, 18] to encourage the model to align the natural language (NL) and programming language (PL) representations. Generally, the MLM is to randomly mask some tokens from the source code and paired documentation and then ask the model to predict the original tokens. In this study, we design two different tasks according to masking the code change or masking the commit message.

**Masked Language Modeling for Code Change (MLM4CC).** In this task, we input the code change and commit message to the model, and mask the code lines from the code change. We focus on the line level rather than the token level in order to keep the integrity format of code diff, following the previous study [33]. Specifically, we randomly sample 15% of the lines in the code diff and mask them, and the model is asked to predict the masked tokens. Note that this task is similar to the Masked Span Prediction (MSP) task proposed in CodeT5 [72] as they both require the model to predict consecutive tokens. The key difference between them is that in MSP, the number of masked tokens is randomly determined, whereas in MLM4CC, we ensure that the masked tokens can form a complete code line. This task helps the model gain general knowledge about the distribution of the code change corpus, and furthermore, when the information from the code change is not enough, the model can refer to the paired commit message to help it make the prediction, which also helps the model build the connection between the NL and PL. From a general perspective, this pre-training task helps the model understand what code change can fulfill the intended functionality (expressed through the commit message). Note that this pre-training task differs from the denoising code diff (DCD) task of CodeReviewer [33], since we also take the commit message as input to build the NL-PL correlations better. In contrast, the previous work only takes the code change as the input and thus can only help the model learn the code change distribution. Formally, the loss can be described as:

$$\mathcal{L}_{MLM4CC}(\theta) = \sum_{t=1}^{k} -logP_\theta(c_t|c^{mask}, m, c_{<t})$$

where $c^{mask}$ is the masked code diff, $m$ is the commit message, $k$ denotes the number of masked code diff tokens, and $c_{<t}$ is the token sequence predicted for the masked code diff so far.

**Masked Language Modeling for Commit Message (MLM4CM).** In this task, we input the code change and commit message to the model, and mask the commit message tokens. Specifically, we randomly sample 15% of the tokens from the commit message which are supposed to be predicted by the model, and then we replace them with the [MASK] token 80% of the time, with a random token 10% of the time, and keep them unchanged 10% of the time, following existing study [9]. This task helps the model gain general knowledge about the distribution of the commit message corpus. Furthermore, when the information from the commit message is not enough, the model can refer to the corresponding code change to help it make the prediction, which again helps the model build the connection between the NL and PL. From a general perspective, this pre-training task helps the model understand the semantic information of the corresponding code change. Similarly, this pre-training task differs from the denoising review comment (DRC) of CodeReviewer, since DRC still merely learns the distribution of the review comments while does not build the NL-PL connection. Formally, the loss of this task can be described as:

$$\mathcal{L}_{MLM4CM}(\theta) = \sum_{t=1}^{k} -logP_\theta(m_t|m^{mask}, c, m_{<t})$$

where $m^{mask}$ is the masked commit message, $c$ is the code diff, $k$ denotes the number of tokens in the masked commit message, and $m_{<t}$ is the token sequence predicted for the masked commit message so far.

*4.3.2 Code Change-Commit Message Dual Generation.* In the above pre-training tasks, the decoder only predicts discrete masked tokens, while in the generation downstream tasks it needs to generate a fluent NL description or a complete code snippet. To fill the gap between the pre-training and downstream tasks, we design two pre-training tasks to train the model for an NL-PL bidirectional conversion, inspired by CodeT5 [72]. To our best knowledge, we are the first to employ such a dual-generation mode to pre-train code-change-oriented models.

**NL → PL Generation (NL2PL).** In this task, we expect the model to learn how to generate the new code based on the old code and the commit message. We consider that code lines in the code diff beginning with [DEL] and [KEEP] can denote the old code before the code change. Therefore, we mask the added code (i.e., code lines beginning with [ADD]), and then send the code diff (which now denotes the old code) and the commit message into the model. The model is asked to predict the masked contents which denote the added content during the code change. Considering that the lines beginning with [ADD] and [KEEP] denote the new code after the code change, the model thus learns how to generate a code snippet during this task. Note that the difference between this task and MLM4CC is that in this task we train the model to explicitly generate the added code (i.e., the lines beginning with [ADD]), while in MLM4CC the masked contents are randomly

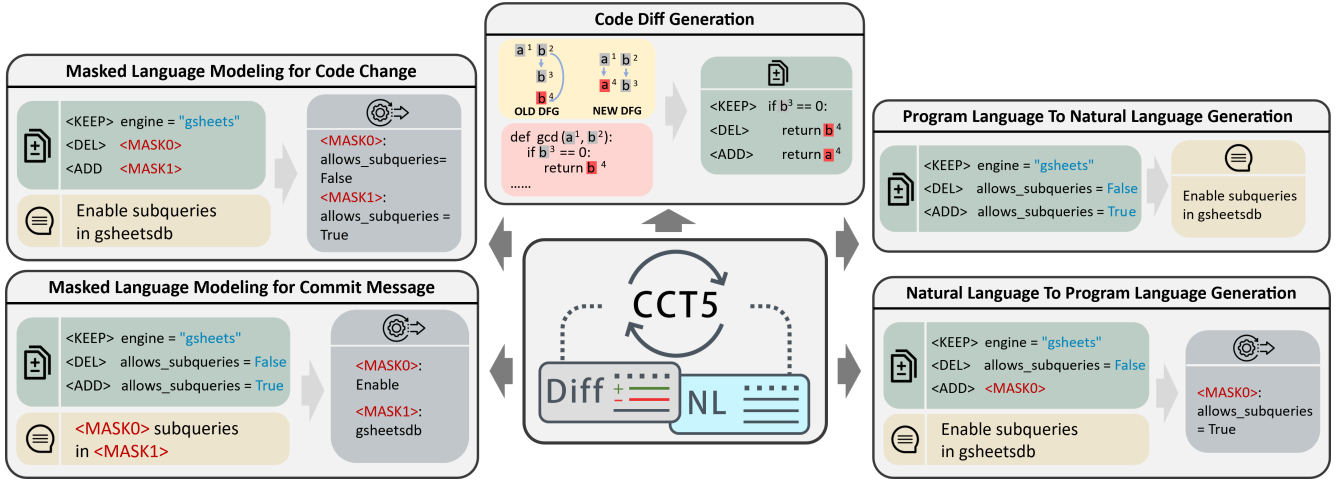

Figure 2: Pre-training tasks of CCT5.

selected. Similar to $\mathcal{L}_{MLM4CC}$, the loss can be described as:

$$\mathcal{L}_{NL2PL}(\theta) = \sum_{t=1}^{k} -logP_\theta(c_t | c^{mask'}, m, c_{<t})$$

where $c^{mask'}$ is the masked code diff, $m$ is the commit message, $k$ denotes the number of masked code diff tokens, and $c_{<t}$ is the token sequence predicted for the masked code diff so far.

**PL → NL Generation (PL2NL).** In this task, we expect the model to learn how to generate the commit message based on the code change. Specifically, we send the code diff into the model and the model is expected to generate the entire commit message, through which the model learns to generate fluent NL descriptions. Formally, the loss can be described as:

$$\mathcal{L}_{PL2NL}(\theta) = \sum_{t=1}^{k} -logP_\theta(m_t | c, m_{<t})$$

where $c$ is the code diff, $k$ denotes the number of tokens in the commit message, and $m_{<t}$ is the token sequence generated so far.

*4.3.3 Code Diff Generation.* The authors of GraphCodeBERT [18] propose that the performance of pre-trained models can be enhanced if considering the code structure during the pre-training. They design two structure-aware pre-training tasks and results show that GraphCodeBERT outperforms CodeBERT [13], which does not involve such pre-training tasks, on a number of downstream tasks. In our study, we also design a structure-aware pre-training task for CCT5 to make it better understand the code change. To our best knowledge, we are the first to involve program structure information when pre-training code-change-oriented models.

Following GraphCodeBERT, we rely on the data flow of the code snippet to provide the semantic-level structure information. Generally, data flow can be considered as a graph in which the data dependency relation (i.e., "where-the-value-comes-from") among different variables is depicted. In the graph, the nodes represent the variables in the code and the edges represent that the two connected variables have dependencies. Data flow is crucial for code understanding since (1) it provides a way to understand the semantics of a variable by concentrating on where its value comes from, rather than focusing on the variable's name, which is sometimes in poor quality (e.g., i and j); and (2) it enables the model to consider the long-range dependency for variables with the same names but occur in distant locations. We refer the readers to the previous study [18] for more details about data flow.

In this **Code Diff Generation (CDG)** task, we send the old data flow, new data flow, and old code into the model, after which the model is asked to generate the corresponding code diff. By doing so, we expect the model could learn to understand the code change based on the data flow change and thus take the program structure into consideration. To adapt the data flow into the acceptable format of the model, we use the variable pair $<Variable_A, Variable_B>$ to represent an edge in the graph that connects two variables, and all the edges are sent into the model sequentially. The inputs are thus represented as { $[CLS]$, $oc_1$, ..., $oc_n$, $[SEP]$, $Variable_A$, $Variable_B$, $[EDGE]$, ..., $Variable_X$, $Variable_Y$, $[SEP]$, $Variable_{A'}$, $Variable_{B'}$, $[EDGE]$, ..., $Variable_{X'}$, $Variable_{Y'}$ }, where $oc_i$ is a token in the old code, $[SEP]$ is a special symbol to split two kinds of data types, $[EDGE]$ is a special symbol to split two edges in the data flow graph, $Variable_I$ denotes a variable in the data flow of the old code, and $Variable_{I'}$ denotes a variable in the data flow of the new code. Formally, the loss can be described as:

$$\mathcal{L}_{CDG}(\theta) = \sum_{t=1}^{k} -logP_\theta(c_t | oc, odf, ndf, c_{<t})$$

where $oc$ is the old code, $odf$ is the data flow before the code change, $ndf$ is the data flow after the code change, $k$ denotes the number of tokens in the code diff, and $c_{<t}$ is the token sequence generated so far.

*4.3.4 Final Loss.* Following the existing studies [6, 13, 38], we treat different pre-training tasks equally. The final loss during the pre-training is calculated as:

$$\min_\theta \mathcal{L}_{Loss}(\theta) = \mathcal{L}_{LML4CC}(\theta) + \mathcal{L}_{LML4CM}(\theta) + \mathcal{L}_{NL2PL}(\theta)$$
$$+ \mathcal{L}_{PL2NL}(\theta) + \mathcal{L}_{CDG}(\theta)$$

## 4.4 Implementation Details

Our model is implemented with the popular deep learning development framework PyTorch.[1] All the experiments are performed on a server with 2 NVIDIA GeForce RTX 4090 GPUs. The learning rate and batch size in the pre-training stage are set to 5e-5 and 32, respectively. When fine-tuning our CCT5 on the downstream tasks which are going to be introduced in the next section, we use a batch size of 32 and a learning rate of 2e-5 for classification tasks (i.e., just-in-time defect prediction) and for the remaining generation tasks, we adopt the same learning rate and batch size as we used during the pre-training stage.

## 5 EXPERIMENTS

The goal of our work is to build a pre-trained model that can be applied to code-change-related downstream tasks. To evaluate the effectiveness of our CCT5, we perform experiments on three different tasks, i.e., commit message generation [41], just-in-time comment update [43], and just-in-time defect prediction [53]. We next elaborate on the three tasks, the baselines, and the results.

## 5.1 Task 1: Commit Message Generation

*5.1.1 Problem Formulation.* It is reported that developers usually do not have enough time to write high-quality commit messages [12]. However, commit messages are of great importance in software maintenance since they describe the intention of the code changes and thus facilitate program comprehension. Therefore, developers could gain considerable benefits from automatic generation of commit messages, making this task a hot topic in software engineering community [23, 27, 41]. In this code-change-related generation task, given the code change, we aim to automatically generate a brief commit message that summarizes its content.

*5.1.2 Baselines.* We select the following state-of-the-art techniques as the baselines of the commit message generation task.

- **NNGen [41].** NNGen is a state-of-the-art retrieval-based commit message generation approach [63]. It first uses "bag-of-words" [58] to represent code changes, after which the code change in the training set with the most similar vector representation to that of the test code change (calculated by the cosine similarity) is identified, whose commit message is reused as the result. According to the recent replication study [63], NNGen is the most effective non-pre-training technique so far for commit message generation.
- **FIRA [11].** FIRA is another state-of-the-art commit message generation approach. It uses a customized graph structure to explicitly depict the code edit operations and then adopts a graph convolution network to represent the code change. Finally, a Transformer layer with the dual copy mechanism is used to generate the message.
- **CodeReviewer [33].** CodeReviewer is a pre-training technique for code changes, but it is pre-trained and evaluated only on code review tasks such as review generation. To assess the effectiveness of this technique on general code change tasks, we fine-tune and evaluate the pre-trained model on our training and test sets.

[1]https://pytorch.org/

Table 2: Performance of different approaches measured by BLEU on the MCMD dataset (in %).

| Dataset | NNGen | CodeT5 | CodeReviewer | CCT5 |
|---|---|---|---|---|
| $MCMD_{java}$ | 17.81 | 17.64 | 18.47 | **20.80** |
| $MCMD_{C\#}$ | 22.92 | 18.76 | 20.36 | **25.53** |
| $MCMD_{c++}$ | 13.69 | 14.41 | 15.94 | **17.64** |
| $MCMD_{python}$ | 16.64 | 18.23 | 17.65 | **21.37** |
| $MCMD_{js}$ | 18.03 | 21.53 | 19.84 | **24.94** |
| Average | 17.82 | 18.11 | 18.45 | **22.06** |

- **CodeT5 [72].** CodeT5 is a state-of-the-art pre-trained model with an encoder-decoder architecture. It achieves the best performance on code-related generation tasks [78]. Technically, it accepts token sequences as inputs so that it can be adapted to code-change-related tasks by feeding it with token sequences of code diffs. Comparing with CodeT5 can better demonstrate the rationale of CCT5 and we also fine-tune and evaluate the pre-trained model on our dataset.

*5.1.3 Dataset & Metrics.* We choose to use the Multi-programming-language Commit Message Dataset (MCMD) [63] as our experiment dataset, which is a recently-released large-scale benchmark for five programming languages including Python, Java, JavaScript, C#, and C++. The total number of commits for each language is 450k and these commits are randomly split in 80-10-10 training/validation/test proportions. However, FIRA cannot be evaluated on this dataset since (1) it currently only supports Java (it uses the *javalang* package to perform program analysis); and (2) it requires complete class files to extract ASTs, whereas the MCMD dataset only contains commit diffs. Therefore, to compare with FIRA, we also evaluate CCT5 on the dataset used in FIRA's evaluation, which is a commonly-used large-scale Java dataset extracted from 1K popular GitHub projects [11, 27, 75]. The dataset contains 75,000 commits as the training set, 8,000 commits as the validation set, and 7,661 commits as the test set.

BLEU is a common metric that measures lexical overlap for evaluating text generation. Among a number of the variants of BLEU, the case insensitive B-Norm has the highest correlation with human evaluations [63]. Therefore, following existing studies [44, 63], we use the B-Norm as our evaluation metric.

*5.1.4 Experiment Details.* Since the commit message generation is a generation task, we use the entire encoder-decoder architecture of CCT5 to perform such a task. The training/validation/test sets are used to fine-tune (train)/validate/evaluate all pre-training (non-pre-training) techniques. We reuse the code as well as the hyper-parameter values released by NNGen, CodeT5, and CodeReviewer to perform this experiment. We reuse the performance of FIRA reported by Dong *et al.* [11] to compare it with CCT5.

*5.1.5 Results.* Results are shown in Table 2. We note that CCT5 consistently outperforms existing approaches concerning all five PLs. On average, the BLEU score of CCT5 on the whole test set is 22.06%, exceeding those of the state-of-the-art pre-training techniques CodeT5 and CodeReviewer, which are 18.11% and 18.45%, by 22% and 20%, respectively. Similarly, CCT5 achieves an increase of 24% when compared with the non-pre-training NNGen (22.06% vs. 17.82%). We further conduct a Wilcoxon signed-rank test [73] between the BLEU scores of CCT5 and the other three baselines on the sub-dataset of each programming language. The results confirm that the difference between the scores of CCT5 and NNGen/CodeT5/CodeReviewer is statistically significant at the confidence level of 95% in all the comparisons. We also note CCT5 achieves comparatively poor performance on C++ language. A potential reason is that this PL is not included in the CodeChangeNet dataset so that the domain knowledge of this PL is not learned during pre-training. Nonetheless, CCT5 still outperforms CodeReviewer, which is pre-trained on data written in C++ language, by around 11% (17.64% vs. 15.94%). Another interesting observation from Table 2 is that CodeReviewer generally outperforms CodeT5 on the MCMD dataset, which also demonstrates the benefits of training data related to code changes.

Table 3: The results of our human evaluation.

| Approach | Adequacy | Conciseness | Expressiveness |
|---|---|---|---|
| NNGen | 2.3 | 2.7 | 4.0 |
| CodeReviewer | 3.0 | 3.6 | 4.4 |
| CodeT5 | 3.2 | 3.7 | 4.3 |
| **CCT5** | **3.4** | **3.9** | **4.6** |

When evaluated on the dataset used by FIRA, CCT5 achieves a BLEU score of 19.30%, which exceeds that of FIRA (17.67%) by 9.2%. The Wilcoxon signed-rank test also shows that the difference between the scores of CCT5 and FIRA is statistically significant at the confidence level of 95%.

*5.1.6  Human Evaluation.* The evaluation metric BLEU can measure the lexical gap between the generated commit messages and the references, but it can hardly reflect the semantic gap. Therefore, we perform a human evaluation to further assess the quality of the commit messages generated by different approaches.

Specifically, we recruit five Ph.D. students who are not co-authors of this paper. All of them have more than 5 years of programming experience and they are majoring in Computer Science. We randomly select 100 commits from the MCMD dataset (20 for each programming language). By applying the four approaches listed in Table 2, we obtained a total of 400 generated commit messages (we exclude FIRA from the human evaluation because it cannot be evaluated on multiple PLs). Following the previous study [63], each participant is asked to rate each generated message from the following three aspects: (1) **Adequacy**, reflects the information richness of the generated messages, (2) **Conciseness**, reflects to what extent extraneous information is included in the generated messages, and (3) **Expressiveness**, reflects to what extent the generated messages are readable and understandable. All these scores are integers ranging from 1 to 5 (1 for poor, 2 for marginal, 3 for acceptable, 4 for good, and 5 for excellent). For each commit, our questionnaire includes the code change, the ground truth commit message, and the commit messages generated by CCT5 as well as the compared techniques. To avoid bias, all four techniques are anonymous in the questionnaire and each participant fills in the questionnaire independently.

Results are shown in Table 3. Overall, CCT5 is better than all the baselines in the three aspects. The average scores of CCT5 for adequacy, conciseness, and expressiveness are 3.4, 3.9, 4.6, respectively. In terms of adequacy and conciseness, the results suggest that CCT5 outperforms existing approaches in capturing the semantic information of code changes since it can generally provide abundant information about the code change while generate little irrelevant information. We also note that all the approaches achieve good performance on expressiveness (the achieved scores are all higher than 4), which means that the generated commit messages are generally readable.

Table 4: An example of the commit message generation task.

| **Code Change**: |
|---|
| ```<br>  except socket.error:<br>    DOCKER_HOST_FROM_CONTAINER = DOCKER_BRIDGE_IP<br>+ if in_docker() and not os.environ.\<br>    get('LAMBDA_REMOTE_DOCKER', '').strip():<br>+   LAMBDA_REMOTE_DOCKER = True<br>  CONFIG_FILE_PATH = \<br>    os.path.join(expanduser("~"), '.localstack')<br>``` |
| **Ground truth**: default to LAMBDA_REMOTE_DOCKER = true if running in Docker( ). |
| **NNGen**: fix host path of Lambda code folder for execution in Docker. |
| **CodeT5**: Merge pull request from localstack/fix/default-LAMBDA_REMOTE_DOCKER. |
| **CodeReviewer**: Code cleanup. |
| **CCT5**: Set LAMBDA_REMOTE_DOCKER = True if running in docker. |

*5.1.7  Case Analysis.* Table 4 shows the commit messages generated by different approaches for a commit of the *localstack* project.[2] We find that the message generated by CCT5 shares the identical semantic information with that written by the developers (i.e., the ground truth): they both express the meaning of setting the value of a field when the program is executing under the docker environment. In contrast, none of the messages generated by other approaches is semantically relevant to the ground truth. For instance, the one retrieved by NNGen is about bug fixing as it contains words such as "fix". Similarly, those produced by CodeT5 and CodeReviewer also fail to capture the semantic of this code change since they explain the intention of the code change as "merge pull request" or "cleanup". Consequently, the participants of the human evaluation rate high scores to the CCT5's result while the other three techniques receive relatively low scores. Specifically, the average scores of CCT5 towards the three aspects are 4.8, 5.0, and 5.0, respectively, while none of the three baselines receives a score higher than 3 with respect to the adequacy. The only participant who gives a score 4 towards the adequacy of CCT5's result explains that she feels there is a subtle semantic gap between the words "default to" and "set". This case shows that existing pre-training techniques, whether designed for code-related tasks (e.g., CodeT5) or trained on data specific to code review tasks (e.g., CodeReviewer), fall short on capturing the semantics of code changes. By designing code-change-oriented pre-training tasks and collecting the pairwise data of code change and commit message, CCT5 can fill the gap well.

## 5.2 Task 2: Just-in-Time Comment Update

*5.2.1  Problem Formulation.* The task of just-in-time (JIT) comment update aims to automatically update code comments according to code changes, so that the comment can keep consistent with the code during the program maintenance and evolution, which will facilitate developers' program comprehension activities [43]. In this code-change-related generation task, the inputs are the original code comment and the code change, while the output is the new code comment.

---
[2]https://github.com/localstack/localstack.

Table 5: Performances of different approaches on the JIT comment update task (in %).

| Approach | Accuracy | GLEU |
| --- | --- | --- |
| Toper | 30.1 | 62.5 |
| CoditT5 | 32.3 | 67.1 |
| CodeReviewer | 23.8 | 63.5 |
| CodeT5 | 22.8 | 64.0 |
| **CCT5** | **36.1** | **70.7** |

*5.2.2 Baselines.* We select the following state-of-the-art techniques as the baselines of the JIT comment update task.

- **Toper [35].** Toper is the state-of-the-art technique regarding the task of JIT comment update, which adopts a combination strategy: given a code change with the old comment, it first uses a classifier to predict if the contents of the comment update can be borrowed from the code change. If so, Toper uses a heuristic based approach [34] to perform the update; otherwise it turns to an deep learning updater. The DL updater embeds the code change through comparing the differences between the old and new code with respect to the ASTs.
- **CoditT5 [79].** CoditT5 is a pre-trained model specially designed for editing tasks. In the pre-training phase, it trains the model to discern edit locations and then identify necessary edit operations. It has also achieved excellent performance on comment updating.
- **CodeReviewer [33] & CodeT5 [13].** These two pre-training techniques have been introduced in the last section. We fine-tune and evaluate them on the datasets of this downstream task.

*5.2.3 Datasets & Metrics.* Following existing studies [34, 35], we use the dataset constructed by Liu *et al.* [43], which contains totally 80,591/8,827/9,204 comment update instances for training/validation/test sets. These 98,622 change instances are all Javadoc comments, which are mined from 1,496 GitHub projects. These comment update instances are further distilled by Lin *et al.* [34] to ensure that the updated contents are not related to changing case of the word or fixing typos.

As the JIT comment update can be considered as an editing task, we use the GLEU [48], a variant of BLEU, as our metric, following a number of studies [42, 79]. We also focus on the Accuracy, which is the percentage of the test samples where correct comments (i.e., those written by the developers) are generated.

*5.2.4 Experiment Details.* Again, we reuse the code and the hyper-parameter values released by CoditT5, CodeT5 and CodeReviewer to perform the experiments. Lin *et al.* [35] have provided the performance of Toper so we reuse the reported results. For the pre-training techniques (i.e., CCT5, CoditT5, CodeT5 and CodeReviewer), the input sequence includes the code diff and the old comment, which are separated by the special symbol [*SEP*].

*5.2.5 Results.* Results are shown in Table 5. We note that comparing with the state-of-the-art approaches, CCT5 not only achieves a higher GLEU value but also generates more correct comments. Specifically, the GLEU value of CCT5 is 70.7%, exceeding that of the most effective existing approach CoditT5 (i.e., 67.1%) by more than 5%. The Wilcoxon signed-rank test shows that the difference between the GLEU scores of CCT5 and Toper/CoditT5/CodeReviewer/CodeT5 is statistically significant at the confidence level of 95%. Also, the Accuracy of CCT5 is 36.1%, which means it can directly output correct comments for more than one third of the test samples. This percentage is also higher than that of CoditT5 (32.3%) by around 12%.

### 5.3 Task 3: Just-in-Time Defect Prediction

*5.3.1 Problem Formulation.* The task of just-in-time (JIT) defect prediction aims to identify defective code changes upon the changes were committed, so that developers can be notified that such changes need further inspection. It is nowadays a key way to track, detect, and fix software defects for large software companies [59, 80]. This code-change-related understanding task is a binary classification: a given code change is predicted as defective or not.

*5.3.2 Baselines.* We select the following state-of-the-art techniques as the baselines of the JIT defect prediction task.

- **JITLine [53].** JITLine fuses features from two perspectives to train a machine learning classifier. First, it uses token-level "bag-of-words" features of code changes. The behind intuition is quite straightforward: code tokens that frequently appear in historical defect-introducing code changes are more likely to trigger defects in the future. Second, it uses commit-level features which have been shown to share a relationship with the likelihood of inducing defects such as the number of added lines and the number of modified files [29, 46].
- **JITFine [50].** JITFine is a hybrid approach which combines the semantic features (SF), mined from the semantic information and syntactic structure hidden in the defective source code, and expert features (EF), which are defined by experts according to their understanding on defect-inducing code changes with their professional knowledge and experience. According to the results of Ni *et al.* [50], such a combination yields the state-of-the-art performance on JIT defect prediction. Note that CCT5 does not involve the expert features. To fairly compare CCT5 with JITFine, we implement a variant of JITFine which only uses the semantic features (denoted as **JITFine - EF**). We also implement a variant of CCT5 where we concatenate the code change vector generated by CCT5 and the the expert feature vector defined in JITFine, and use the combined vector for prediction (denoted as **CCT5 + EF**).
- **CodeReviewer [33] & CodeBERT [13].** As reported by Zend *et al.* [78], CodeBERT is more effective than CodeT5 on JIT defect prediction task so that we use it as the representative pre-trained model here. Note that JITFine obtains the numeric representation of semantic features by sending the code change into the pre-trained CodeBERT model. Hence, the performance of CodeBERT actually represents the effectiveness of JITFine - EF on this task.

*5.3.3 Datasets & Metrics.* Following the existing study, we choose to use the JIT-Defects4J [50] as the evaluation dataset, which is built upon a previous dataset named LLTC4J [19]. Tangled commits are commonly-observed which means there may be several kinds of code changes in a commit such as bug fixing, code refactoring, and new features [20]. Therefore, traditional datasets collected by labelling the commits related to all the changed lines in a bug-fixing commit as defect-inducing, such as the one proposed by McIntosh and Kamei [46], may be inaccurate and bring the noise to the approach evaluation. To address this challenge, Ni et al. rely on the LLTC4J dataset which is manually labelled by 45 experienced participants to determine the role of each changed line in bug-fixing commits, i.e., whether the changed line is to fix bugs, change whitespace, change documentation, etc. Based on the labelling results, Ni et al. collect the bug-introducing commits of those lines identified as "contributing to the bug-fixing" and consider such commits as defect-inducing code changes in the JIT-Defects4J dataset [50]. Such an operation can greatly reduce the scope of defect-inducing candidates and thus alleviate the impact of tangled commits.

Table 6: Performances of different approaches on the JIT-Defects4J dataset.

| Approach | F1 | AUC |
| --- | --- | --- |
| JITLine | 0.261 | 0.802 |
| JITFine | 0.431 | 0.881 |
| CodeBERT (JITFine - EF) | 0.375 | 0.856 |
| CodeReviewer | 0.341 | 0.802 |
| CCT5 | 0.451 | 0.871 |
| CCT5 + EF | **0.472** | **0.882** |

The JIT-Defects4J dataset contains a total of 27,319 code changes from 21 open-source large-scale Java projects, among which 2,332 code changes are defect-inducing and the other 24,987 code changes are clean, making the bug ratio being around 8.5%. The training, validation, and test sets are split by timestamp. Specifically, the top 80% of commits in each project are used for training (60% as the training set and 20% for validation), while the rest 20% of commits in each project are treated as test data.

Following existing studies [22, 23], we use the F1-score and the Area Under the receiver operator characteristics Curve (AUC) as the metrics of a model's discriminatory power, i.e., its ability to differentiate between defective and non-defective code changes. When computing the F1-score, the code change which induces a defect is treated as the positive class.

5.3.4 Experiment Details. Since the JIT defect prediction is an understanding task, we only use the pre-trained encoder of CCT5 to perform the experiment. Specifically, the representation of the special token [CLS] in the last layer of the encoder is used as the representation of the code change, and it is fed into a linear classifier to produce the prediction. To perform this experiment, we reuse the open sourced implementations of CodeReviewer and CodeBERT and connect their encoders to a linear classifier, whose architecture is identical to that of CCT5, to output the final results. Ni et al. [50] have evaluated JITLine and JITFine under the same experiment setting, so we directly reuse their results for comparison.

5.3.5 Results. Results are demonstrated in Table 6. We observe that the vanilla CCT5 has already achieved an F1-score higher than those of the existing techniques. Specifically, the F1-score of CCT5 is 0.451, outperforming the best-performing pre-training technique, i.e., CodeBERT (0.375), and the best-performing non-pre-training technique, i.e., JITLine (0.261), by around 20% and 73%, respectively. It also outperforms the state-of-the-art hybrid approach JITFine by around 5% (0.451 vs. 0.431). When incorporating the expert features, CCT5 achieves the highest F1-score and AUC score among all the techniques. Specifically, the F1-score of CCT5 increases from 0.451 to 0.472 after using the extra features from JITFine. Such results indicate that CCT5 is extensible: its performance holds the potential for enhancement after involving more useful features.

Table 7: Performances of CCT5 on code review tasks.

| Approach | Code Change Quality Estimation | | Review Generation |
| --- | --- | --- | --- |
| | F1 | AUC | BLEU (in %) |
| CodeReviewer | 0.715 | 0.742 | 5.32 |
| CCT5 | **0.776** | **0.860** | **6.30** |

## 6 DISCUSSION

### 6.1 Can CCT5 Generalize Well to Code Review Tasks?

In the last section, we have demonstrated that CCT5 outperforms CodeReviewer in three popular code-change-related tasks. This is within our expectation since CodeReviewer is pre-trained only on data collected from code review tasks, whose characteristics may be different from those of our evaluation datasets. In this section, we evaluate CCT5 on code review tasks, aiming at better investigating the generalizability of CCT5.

We focus on one understanding task (i.e., code change quality estimation) and one generation task (i.e., code review generation). Code change quality estimation task is to predict whether a code change is high-quality and can be already accepted. This is a binary classification: a code change can be predicted as high-quality or not. Based on the prediction results, code reviewers can give questionable code changes a higher priority to review and pay more attention to them, saving time and effort. Code review generation task is to automatically generate a review comment based on the given code change, make the code reviewers free from writing comments manually, and thus lighten the burden of reviewers.

The evaluation datasets have already been provided by Li et al. [33]. Specifically, they crawled pull requests and reviewed comments of 1,176 popular GitHub projects written in a total of nine programming languages. For the code change quality estimation task, all code changes with review comments are considered as low-quality while those without comments are labelled as correct. For the code review generation task, only the code changes with review comments are used for evaluation. Specifically, the training/validation/test sets for code change quality estimation contain ∼266K/∼31K/∼31K code changes respectively, and the training/validation/test sets for code review generation contain ∼118K/∼10K/∼10K code changes respectively.

Following the evaluation setting in the last section, we use F1-score and AUC as the metrics for code change quality estimation (note that when performing the calculations, low-quality code changes, i.e., those need review comments are treated as the positive class), and we use the BLEU-4 score for code review generation.

Results are shown in Table 7. We note that CCT5 outperforms CodeReviewer on both tasks. Specifically, the F1-score of CCT5 on the code change quality estimation task reaches 0.776, exceeding that of CodeReviewer (i.e., 0.715) by nearly 9%. Similarly, the BLEU score of CCT5 on review generation is 6.30%, exceeding that of CodeReviewer (i.e., 5.32%) by 18%.

**Case Analysis.** One example is shown in Table 8. In the depicted code change of a Go program, the return value of the function call `os.RemoveAll()` is passed to a local variable. However,

Table 8: An example of the code review generation task.

| **Code Change**: |
| --- |
| ```
    if err != nil {
        t.Fatal(err)
    }
-   return b, func() {}
+   return b, func() { a = os.RemoveAll(dir) }
``` |
| **Ground truth**: The body can just be 'os.RemoveAll(dir)'. |
| **CodeReviewer**: I don't think we need to use 'os.RemoveAll'. |
| **CCT5**: Shouldn't this be 'os.RemoveAll(dir)'? |
| **An Example in the Pre-training dataset**: |
| ```
    return func() {
+       os.RemoveAll(dir)
    }, nil
``` |

Table 9: Results of the ablation study (in %).

| Model | −MLM4CC | −MLM4CM | −NL2PL | −PL2NL | −CDG | CCT5 |
| --- | --- | --- | --- | --- | --- | --- |
| GLEU | 68.5 | 68.1 | 68.6 | 67.7 | 68.0 | **70.7** |

this function call actually has no return value. Therefore, CCT5 and the developer suggest only performing the function call without passing the return value. In contrast, the review generated by CodeReviewer suggests completely removing this function call, which is inappropriate. By investigating the pre-training dataset, we find some similar code changes, one of which is shown in the second part of the table. Via learning the data flow relation through the CDG pre-training task, CCT5 could obtain the knowledge that there should not be any data flow relation between the argument of this function call and other variables. When evaluated on the depicted code change, CCT5 would understand that the code is incorrect since there is a variable that contains data flow relation with the argument of this function call. That could be the reason why CCT5 generates a semantically-correct review comment. This case demonstrates that CCT5 outperforms CodeReviewer on code review tasks probably because it can better capture the data flow dependency of the program.

### 6.2 The Contribution of Each Standalone Pre-Training Task

We have demonstrated that the well-designed pre-training tasks help CCT5 achieve the state-of-the-art performance on diverse code-change-related tasks. We further investigate the contribution of each pre-training task. Specifically, we pre-train five variants of CCT5, each of which is trained with one pre-training task removed, and then evaluate them on downstream tasks.

Table 9 shows the results on the JIT comment update task (results on other tasks demonstrate similar trends and we only illustrate the results on this task due to space limitation). Overall, all pre-training tasks contribute to the performance of CCT5. Specifically, we note that the effectiveness of CCT5 decreases the most (3%) if the PL2NL pre-training task is excluded. This is within our expectation since a number of downstream tasks require the ability to generate fluent natural language descriptions (e.g., commit message generation and comment update). The GLEU score also decreases by 2.7% without the CDG task, which indicates that making the model structure-aware is critical for an effective model.

### 6.3 Threats to Validity

**External Threats.** Beyond the three widely-studied downstream tasks evaluated in this paper, there are a number of other code-change-related tasks such as automated patch correctness assessment [36, 70], bug-fixing patch identification [23, 24], and untangling commits [10, 51]. We leave evaluating the effectiveness of CCT5 on such tasks as our future work. In addition, the emergence of large language models (LLMs), such as ChatGPT, is reshaping our research landscape. It would be valuable to investigate the effectiveness of such LLMs on code-change-related tasks with zero-shot or few-shot learning. We leave conducting a thorough investigation as our future work.

**Internal Threats.** The success of deep learning techniques relies heavily on the quality of the training data. In this study, we only adopt simple heuristics to filter low-quality data when building the pre-training dataset, CodeChangeNet. It is thus possible that the pre-training dataset contains some noisy data whose commit message does not accurately match with the code change. As revealed by the previous study [62], such noisy data could negatively affect the effectiveness of CCT5. Our CCT5 could also be undertrained because the CodeChangeNet dataset only contains data from six programming languages and is thus of limited scale. At present, other popular programming languages, such as C and Scala, are not being considered in this study. Therefore, future research efforts could be devoted to build a larger and higher quality pre-training dataset. To compare CCT5 with the existing approaches, we reproduce a number of state-of-the-art techniques. To ensure the reliability of our experiment results, we carefully reuse the code as well as the hyper-parameter values released by the previous study. Besides, to avoid the data leakage, we carefully remove all projects that are used to build the dataset of the five downstream tasks from our pre-training dataset, i.e., CodeChangeNet. We also leave performing human evaluation on the comment update and review generation tasks as our future works.

## 7 SUMMARY

Developers usually spend much effort on software maintenance, but the literature needs an effective way to help them deal with tasks derived from code changes. To fill this gap, we propose to build a code-change-oriented pre-trained model which can both leverage the domain knowledge hidden in the large-amount unlabelled data and ensure the learned knowledge is adequately exploited during fine-tuning on code-change-related tasks. To achieve so, we collect a large-scale dataset for pre-training and design five different tasks to equip the model with domain knowledge about code changes. Our evaluation shows that the pre-trained model, CCT5, achieves state-of-the-art performance on totally five downstream tasks.

All data in this study are publicly available at: **https://github.com/Ringbo/CCT5**.

## ACKNOWLEDGMENTS

This work is supported by the National Natural Science Foundation of China No.61932021 and No.62202420. Zhongxin Liu gratefully acknowledges the support of Zhejiang University Education Foundation Qizhen Scholar Foundation.